\documentclass[]{spie}  
\usepackage[dvips]{graphicx}

\title{SNAP Near Infrared Detectors} 

\author{G.~Tarl\'e\supit{a}, C.~Akerlof\supit{a},
  G.~Aldering\supit{b}, R.~Amanullah\supit{c}, P.~Astier\supit{d},
  E.~Barrelet\supit{d},
  C.~Bebek\supit{b},\\ L.~Bergstr\"{o}m\supit{c}, 
  J.~Bercovitz\supit{b}, G.~Bernstein\supit{e}, 
  M.~Bester\supit{f}, A.~Bonissent\supit{g}, C.~Bower\supit{h},\\
  M.~Brown\supit{a}, W.~Carithers\supit{b}, E.~Commins\supit{f}, 
  C.~Day\supit{b},
  S.~Deustua\supit{i}, R.~DiGennaro\supit{b}, A.~Ealet\supit{g},\\
  R.~Ellis\supit{j}, M.~Eriksson\supit{c}, A.~Fruchter\supit{k},
  J-F.~Genat\supit{d}, G.~Goldhaber\supit{f}, A.~Goobar\supit{c},
  D.~Groom\supit{b},\\
  S.~Harris\supit{f}, P.~Harvey\supit{f}, H.~Heetderks\supit{f},
  S.~Holland\supit{b}, D.~Huterer\supit{l}, A.~Karcher\supit{b},
  A.~Kim\supit{b},\\ W.~Kolbe\supit{b}, B.~Krieger\supit{b},
  R.~Lafever\supit{b},
  J.~Lamoureux\supit{b}, M.~Lampton\supit{f}, M.~Levi\supit{b},
  D.~Levin\supit{a},\\ E.~Linder\supit{b},  S.~Loken\supit{b},
  R.~Malina\supit{m}, R.~Massey\supit{n}, R.~Miquel\supit{b}, 
  T.~McKay\supit{a},
  S.~McKee\supit{a},\\
  E.~M\"{o}rtsell\supit{c}, N.~Mostek\supit{h}, S.~Mufson\supit{h},
  J.~Musser\supit{h}, P.~Nugent\supit{b}, H.~Oluseyi\supit{b},
  R.~Pain\supit{d},\\ N.~Palaio\supit{b}, D.~Pankow\supit{f}, 
  S.~Perlmutter\supit{b},
  R.~Pratt\supit{f}, E.~Prieto\supit{m}, A.~Refregier\supit{n},
  J.~Rhodes\supit{o},\\ K.~Robinson\supit{b}, N.~Roe\supit{b},
  M.~Schubnell\supit{a}, M.~Sholl\supit{f}, G.~Smadja\supit{p},
  G.~Smoot\supit{f},\\ A.~Spadafora\supit{b}, 
  A.~Tomasch\supit{a}, D.~Vincent\supit{d}, H.~von der Lippe\supit{b}, 
  J.~Walder\supit{b} and G.~Wang\supit{b}
\skiplinehalf
\supit{a}University of Michigan, Ann Arbor MI, USA\\
\supit{b}Lawrence Berkeley National Laboratory, Berkeley CA, USA\\
\supit{c}University of Stockholm, Stockholm, Sweden\\
\supit{d}CNRS/IN2P3/LPNHE, Paris, France\\
\supit{e}University of Pennsylvania, Philadelphia PA, USA\\
\supit{f}University of California, Berkeley CA, USA\\
\supit{g}CNRS/IN2P3/CPPM, Marseille, France\\
\supit{h}Indiana University, Bloomington IN, USA\\
\supit{i}American Astronomical Society, Washington DC, USA\\
\supit{j}California Institute of Technology, Pasadena CA, USA\\
\supit{k}Space Telescope Science Institute, Baltimore MD, USA\\
\supit{l}Case Western Reserve University, Cleveland OH, USA\\
\supit{m}CNRS/INSU/LAM, Marseille, France\\
\supit{n}Cambridge University, Cambridge, UK\\
\supit{o}Goddard Space Flight Center, Greenbelt MD, USA\\
\supit{p}CNRS/IN2P3/IPNL, Lyon, France
}

\authorinfo{Further author information: (Send correspondence to
  G.T.)\\G.T.: E-mail: gtarle@umich.edu, Telephone: 1 734 763 1489}

\begin{document} 
  \maketitle 

\begin{abstract}
The SuperNova/Acceleration Probe (SNAP) will measure precisely the
cosmological
expansion history over both the acceleration and deceleration epochs
and thereby constrain the nature of the dark energy that dominates our
universe today.
The SNAP focal plane contains equal areas of optical CCDs and NIR
sensors and an integral field spectrograph.
Having over 150 million pixels and a field-of-view of 0.34 square
degrees, the SNAP NIR system will be the largest yet constructed. With
sensitivity in the range 0.9--1.7 $\mu$m, it will detect
Type Ia supernov{\ae} between $z = 1$ and 1.7 and will provide
follow-up precision photometry for all supernov{\ae}.
HgCdTe technology, with a cut-off tuned to 1.7 $\mu$m,
will permit passive cooling at 140 K while maintaining noise
below zodiacal levels.
By dithering to remove the effects of intrapixel variations and by
careful attention to other instrumental effects, we expect to control
relative photometric accuracy below a few hundredths of a magnitude.
Because SNAP continuously revisits the same fields we will be able to
achieve outstanding statistical precision on the photometry of
reference stars in these fields, allowing precise monitoring of
our detectors. The capabilities of the NIR system for
broadening the science reach of SNAP are discussed.  
\end{abstract}



\section{INTRODUCTION}
\label{sect:intro}  

Near infrared observations from space are essential to identify the nature of
the dark energy responsible for the acceleration of the universe that
we see today.
For redshifts of order one or larger  the universe was dominated by
matter and underwent deceleration, while for redshifts of order one or
less the universe has been dominated by dark energy resulting in
acceleration.
Whereas measurements at low redshift provide the best determination of
the dark energy content of our universe today, measurements at high
redshift provide the best opportunity to distinguish among
dark energy models and to control systematic errors. The
SuperNova/Acceleration Probe (SNAP)\cite{Aldering} will obtain precision
calibrated light curves and spectra for over 2500 Type Ia supernov{\ae} at
redshifts from 0.1 to 1.7 to determine the nature of the dark energy.


For supernov{\ae} at redshifts above one, the rest-frame visible emission
at 500 nm is shifted beyond the long-wavelength
cutoff of optical CCD detectors and into the near infrared.
Supernova (SN) photometry must be performed at a common rest-frame (blue)
wavelength if the supernov{\ae} are to be used as standard candles;
hence, NIR imaging at wavelengths of 1.0--1.7 $\mu$m is a critical
aspect of the mission. Indeed the NIR system on SNAP provides {\it
 all} the restframe optical photometry above $z=1.2$. NIR photometry is
also essential to the diagnosis and correction of potential systematic
errors caused by gray dust, SN evolution and SN magnification.
As an integral part of the SNAP observation program, NIR observations
will allow photometric redshift determination for all SNAP galaxies, 
providing a trigger
for rejecting high-$z$ Type II supernov{\ae}.
The unique wide-field SNAP NIR system also permits a significant 
expansion of the
science reach of SNAP beyond the core mission.

Extensive mission simulations have
shown that a wide-field NIR camera integrated with
optical CCDs into a single focal plane provides the optimal configuration
to extract the most cosmological information. In
particular, the measurement of broadband colors for every supernova
provides the necessary leverage to determine both the equation of
state $w \equiv p/\rho$ of the universe and its time variation with sufficient
precision to constrain dark energy models.
As presently conceived, the NIR detectors for SNAP constitute fully half
of the focal plane with every supernova observed in every color from
350 to 1700 nm.

\section{Identifying the Nature of Dark Energy} 
\label{sect:dark_energy}

SNAP will use measurements of Type Ia supernov{\ae} as a probe of the
nature of dark energy.
The essential SNe measurement for this purpose is a precise comparison
of luminosity distance to redshift from $z=0$ to $z=1.7$.

It has been shown that Type Ia supernov{\ae} have very uniform peak
$B$-band brightness, but only when they are corrected for a stretch
factor which describes the relation between absolute brightness and
explosion duration\cite{Perlmutter}. To fully standardize the peak
brightness of a supernova, a variety of additional observations must
be made. The color throughout the light curve provides constraints on
both host galaxy and Milky Way extinction.
The nature of the host galaxy and the location of the supernova
within it may provide important additional information. Finally, a
spectrum of the supernova obtained near maximum light will allow
identification of the explosion as a Type Ia as well as of secondary
supernova characteristics.

To construct the desired SNe Hubble diagram, restframe $B$ and $V$
lightcurves, restframe $B$ and $V$ galaxy images, and an accurate host
redshift are required. 
At $z=1.7$, restframe $B$ and $V$ have shifted beyond
the sensitivity of CCDs, well into the NIR.
Observations in the NIR thus are crucial to the core project of SNAP.
Without them SNAP cannot directly compare restframe peak brightnesses
of SNe across a wide range of redshifts.

   \begin{figure} [b]
{\begin{minipage}{3.2in} {
   \includegraphics [width=3.0in] {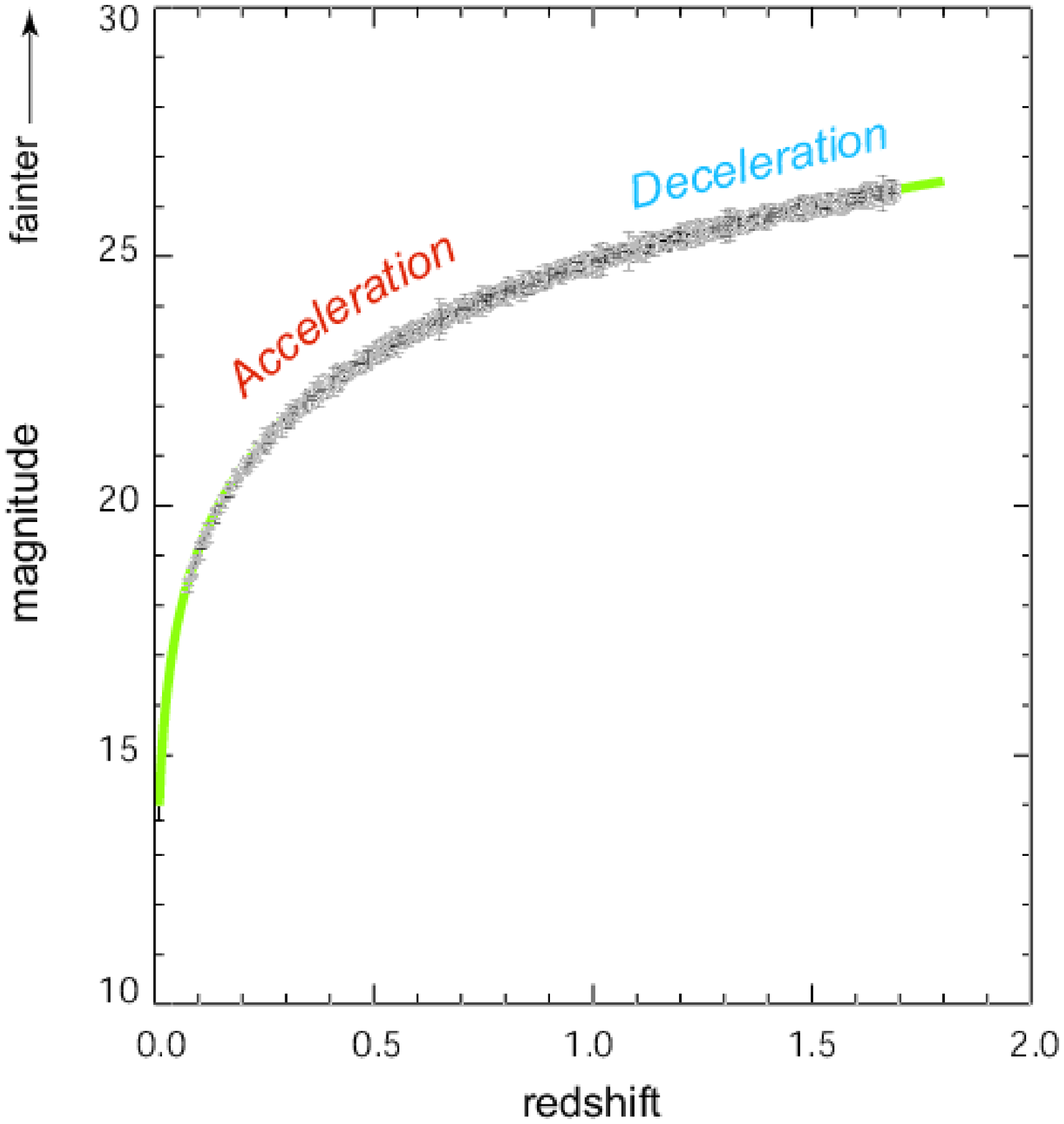}
   \caption[example] 
   { \label{fig:hubble_diagram} 
Simulation of the Hubble diagram that will be obtained by SNAP
containing over 2000 Type Ia supernov{\ae} spanning both the acceleration
and deceleration epochs of the universe.}
}
\end{minipage}
\begin{minipage} {.3in}
{}
\end{minipage}
\begin{minipage} {3.2in}{
   \includegraphics [width=2.7in] {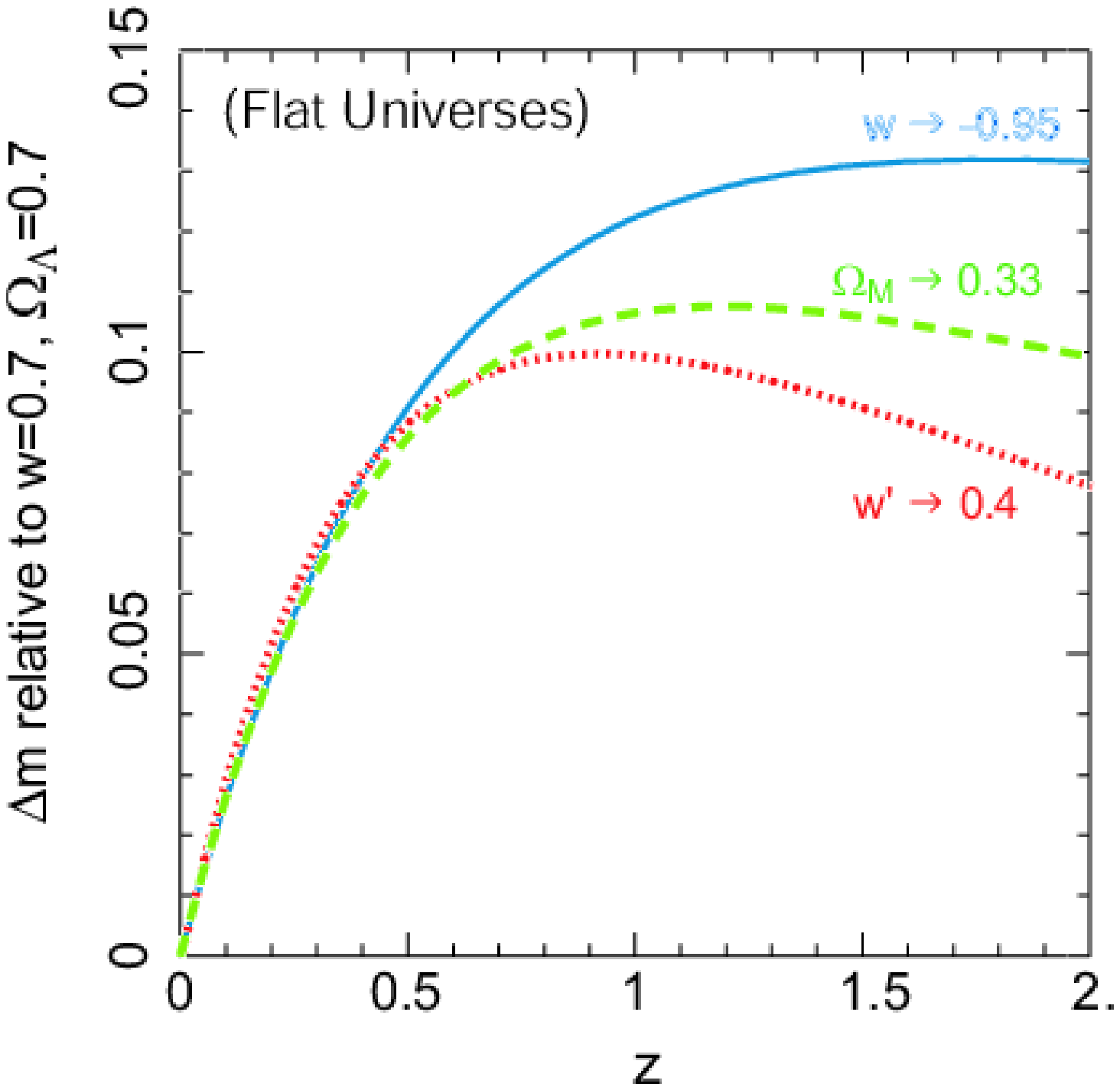}
   \caption[example] 
   { \label{fig:de_models} 
The shift in magnitude, relative to a cosmology with
$\Omega_\Lambda=0.7$ and $w=-0.7$ for three different flat cosmological
models. Note that high-redshift measurements in the near infrared are
essential to distinguish among models of dark energy causing the
acceleration of the universe.}
}
\end{minipage}}
\end{figure}


At low redshift two independent
measurements\cite{Perlmutter,Riess} have
determined that the fraction of the critical density consisting of
dark energy, $\Omega_\Lambda$, is nonzero and possibly of order
unity. Measurement of small scale fluctuations in the cosmic microwave
background radiation subsequently have determined that the universe is nearly
flat\cite{Jaffe}.  Observations of galaxy 
clustering\cite{Bahcall}
have shown that the fraction of the critical density consisting of
matter is $\Omega_M$ = 0.3, consistent with the
results obtained from the supernova measurements.  This implies then
that $\Omega_\Lambda \sim {2\over
3}$ and $\Omega_M \sim {1\over 3}$. 

SNAP will obtain a
magnitude-redshift (Hubble) relation containing over 2000 Type Ia
supernov{\ae}, extending to $z = 1.7$ (see Fig.~\ref{fig:hubble_diagram}). 
Cosmological parameters
can be extracted from this Hubble diagram, including a precise value
for the dark energy equation of state 
(predicted to be $w = -1$ for a
cosmological constant) and constraints on its generically expected
time variation $w' \equiv dw/dz$. Figure \ref{fig:de_models} shows the
shift in magnitude relative to a universe with $\Omega_\Lambda=0.7$ and
$w=-0.7$ for flat universes with slightly different values of $w, w'$,
and $\Omega_M$ varied one at a time.
As can be seen, the greatest sensitivity to cosmological parameters is
obtained with high redshift measurements in the NIR. 

NIR observations will also allow us to control various systematic errors.
While restframe $B$ and $V$ photometry provide some
constraint on the impact of ordinary dust on SNe measurements, they may
not provide sufficient information to constrain ``gray''
dust. Truly gray dust is undetectable by color measurement, but physically
motivated models of dust are actually only ``grayer'' dust; they still
affect light of different wavelengths differently. The best
opportunity to detect such gray dust is to observe each supernova
over the broadest possible wavelength range. NIR observations extend
the wavelength range over which each supernova will be imaged by more
than a factor of two, providing a better chance to constrain a wider 
variety of dust models.

Supernova evolution is an important systematic that is controlled primarily
through spectroscopic observations. Nearby supernov{\ae} span the entire
range of metallicity and other SN parameters that will be observed for
more distant supernov{\ae}. It is unlikely that population changes will
go unnoticed by SNAP. However, Type Ia SNe simulations
have shown that restframe $V-R$ photometry may be less sensitive to
changes in metallicity than $B-V$ and thus NIR measurements 
provide an important
cross check on this systematic. Restframe R measurements are possible
up to $z=1.3$ with the SNAP NIR detectors. 

A significant factor in operation of the SNAP satellite is the
supernova trigger. While the full survey area will be regularly imaged
in all colors, so that every supernova will have full photometric coverage, 
Type Ia supernov{\ae} must be identified at the earliest possible time
for targeting of spectroscopic observations. This trigger can be
substantially enhanced by prior knowledge of host galaxy redshifts.
These can be estimated quite accurately from high S/N
measurements of broadband colors\cite{Connolly}. This is especially simple when
the available colors span the restframe 4000~\AA\ break. Since this
break appears at 0.8 $\mu$m at redshift 1, and 1.2 $\mu$m at redshift
2, NIR information is essential for accurate measurement of photometric
redshifts in this range. With broadband colors from 350 to 1700 nm,
SNAP can readily measure photometric redshifts for 
galaxies from $z=0$ to $z=3.2$.
Knowing the host
galaxy redshift in advance helps to eliminate Type II supernov{\ae} and
other transients from the
sample targeted for resource-intensive spectroscopic observations.

To complement its supernova cosmology observations, SNAP will conduct a 
wider area weak lensing survey. These weak lensing observations provide
important independent measurements and complementary determinations 
of the dark matter and dark energy
content of the universe. They will substantially enhance SNAP's ability
to constrain the nature of dark energy\cite{Huterer}. SNAP weak 
lensing observations benefit enormously from the high spatial
resolution, the accurate photometric redshifts, and the very high surface 
density of resolved galaxies available in these deep observations. SNAP NIR
observations enable the photometric redshift measurements which 
allow SNAP lensing to track the evolution of the matter power spectrum.
This substantially tightens the lensing constraints on the equation of state.

\section{SECONDARY SCIENCE MADE POSSIBLE BY NIR OBSERVATIONS}
\label{sect:sec_science}

NIR observations made from the ground are hindered by the brightness and
opacity of the night sky. As a result, space-based NIR observations augment
even the most powerful ground-based instruments in important
ways. We outline here a few areas in which SNAP NIR survey observations
will impact our understanding of the universe. SNAP will conduct two
primary surveys, a $\sim15$ square degree ultradeep ($m_{AB} \sim 30$
for point sources)
supernova survey, and a $\sim 300$ square degree shallower 
($m_{AB} \sim 27.8$ for point sources) weak lensing survey.


{\it Galaxy evolution and clustering:} Within the 15 square degree supernova
survey area, SNAP will make accurate photometric redshift measurements 
for at least 
$5\times10^{7}$ galaxies from redshift 0 to 3.5. This data set, which will include 
morphological information for every object, will provide a unique opportunity 
to study the evolution of galaxies through more than 90\% of the age of the 
universe. The utility of this data set for galaxy evolution is hinted at by 
the flood of galaxy evolution papers based on the Hubble Deep Fields.

{\it Identification of high redshift galaxy clusters:} Galaxy clusters, the 
most massive bound objects in the universe, provide important probes of our 
understanding of structure formation. Constraining their formation and 
evolution is an important observational goal for the coming decade. Recent 
advances have overcome earlier limitations of 
optically selected cluster samples, essentially by using photometric 
redshift information to eliminate projection effects. The SNAP surveys will
provide detailed information on roughly 15,000 galaxy clusters with masses
above $5\times10^{13}$ M$_\odot$.

{\it A census of $R$-, $I$-, $z$-, and $J$-band dropout galaxies:} Photometric
redshifts can be estimated from the 4000~\AA\ break for galaxies
out to about $z = 3.2$.  For galaxies at still higher redshift, the simplest
indicator is the Lyman break. For SNAP, the Lyman break enters the optical
imager around redshift 3. In principle it can be followed using SNAP
data beyond redshift 10, allowing identification of extremely high
redshift galaxies.

\begin{figure} [t]
   \begin{center}
   \begin{tabular}{c}
   \includegraphics [height=3in] {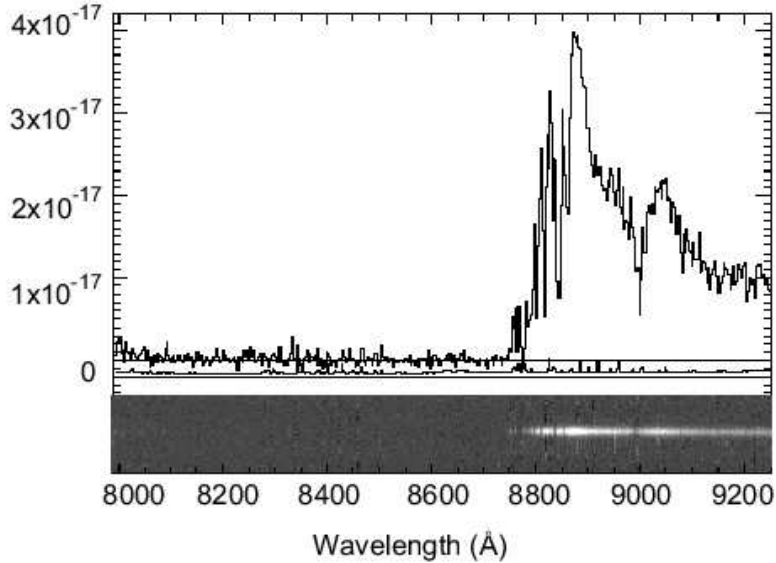}
   \end{tabular}
   \end{center}
   \caption[example] 
   { \label{fig:sdss_z6} 
This figure, from Ref.~\citenum{pen02},
shows a VLT/FORS2 spectrum of the $z = 6.28$ quasar SDSS J1030+0524 in
the observed frame. The bottom panel shows a gray-scale representation
of the sky-subtracted two-dimensional spectrum plotted on the same
wavelength scale. Note the apparent Gunn-Peterson trough; a complete
absence of flux from 8450 to 8710~\AA.}
   \end{figure}

{\it Mapping the quasar luminosity function to $z = 10$:} Quasars are identified 
in multi-color imaging surveys by their non-stellar colors. This method has 
been shown by the Sloan Digital Sky Survey to be extremely effective at 
identifying  quasars to redshift 6 and beyond. SDSS quasar discovery is 
limited to redshift 6 by the CCD sensitivity cutoff at 1.0 $\mu$m\cite{pen02} 
(see Fig.~\ref{fig:sdss_z6}). The most distant SDSS
quasar, at redshift 6.28, has a $z$-band magnitude of 20. By probing to
wavelengths 1.7 times greater, and to depths 9 magnitudes fainter,
SNAP will be able to detect quasars beyond redshift 10, and to probe
the quasar luminosity function to 100 times fainter than the
brightest quasars.

{\it Studies of GRBs to very high redshift:} Current evidence suggests
that gamma-ray bursts are associated with the collapse of massive
stars which live short lives and die where they are born. As a result,
GRBs may trace the cosmic star formation rate.  If so, there should be
GRBs essentially coincident with the epoch of formation of the first
stars. The most distant GRB known occurred at redshift of 3.4. SNAP
will be able to identify GRB afterglows, and the orphan afterglows
predicted by some models of beaming in GRBs to $z = 10$. Such orphan
afterglows may even be detected as backgrounds to the SNe search.

{\it Probing the structure of reionization:} The universe became
neutral at the time of recombination, around $z = 1000$, and the thermal
radiation from that epoch travels to us undisturbed as the cosmic
microwave background radiation. The lack of a Gunn-Peterson effect in
the spectra of most quasars demonstrates that the universe was
reionized at some time between $z = 1000$ and $z = 6$. The source of the
ionizing radiation is the subject of substantial speculation.
The recent discovery of an apparent Gunn-Peterson trough in the most
distant $z > 6$ SDSS quasar spectra may provide the first glimpse of the
epoch of reionization\cite{Becker}.  By identifying many quasars and galaxies to
$z=10$, SNAP will set the stage for mapping the epoch of reionization in
unprecedented detail. In combination with ground based and NGST
spectroscopy, it will enable measurements of the proximity effect and
studies of the spatial structure of reionization.

\begin{figure} [h]
   \begin{center}
   \begin{tabular}{c}
   \includegraphics [height=4in] {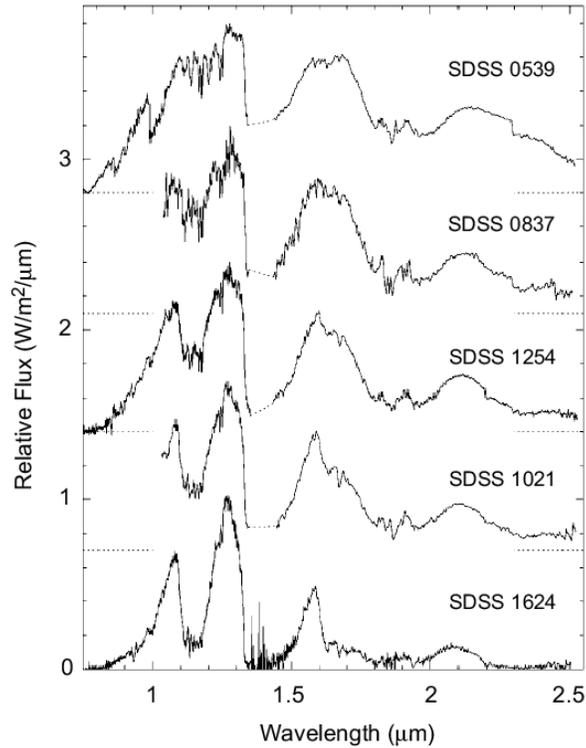}
   \end{tabular}
   \end{center}
   \caption[example] 
   { \label{fig:l_and_t_dwarves} 
This figure, from Ref.~\citenum{leg00}, shows optical/NIR
spectra for five stars ranging from a late L dwarf at the top through
a series of T dwarves. The nearly complete absence of optical flux from
these very cool stars, along with their strong NIR emission, is apparent.}
   \end{figure}

{\it Discovering the coolest nearby objects:} Most of the applications
above stress the ability of SNAP NIR observations to probe restframe
UV and optical emission in objects at high redshift. These NIR
observations can also be used to probe very sensitively the restframe
NIR emission from nearby objects. Of particular interest would be a
census of low-mass L and T stars and brown dwarfs throughout the Milky
Way disk\cite{leg00} (see Fig.~\ref{fig:l_and_t_dwarves}).

{\it Gravitational lensing measurements:} The high spatial resolution of
SNAP NIR observations will enable the discovery of a large number of
new strong lenses. The NIR observations, which are much less sensitive
to dust extinction within the lens galaxy, are especially important in
this regard. In weak lensing measurements, SNAP spatial resolution and
NIR sensitivity will allow the use of a huge number of faint, high
redshift background galaxies. Because these source galaxies must be
resolved to be useful for lensing measurements, and because galaxies
fainter than R=25 have half-light radii less than 0.2$''$, these
measurements are impossible from the ground. With these galaxies, 
it will be possible to extend weak
lensing studies to lower mass objects, and to study lens objects
beyond $z = 1$.

{\it Discovery of outer solar system objects:} SNAP time series data
will provide an excellent probe of faint, red objects in the Kuiper
belt and beyond.  A 2-3 month SNAP survey would detect 10--50,000
Kuiper belt objects down to the size of the Halley's nucleus.

{\it Identification of targets for spectroscopic instruments:} In the
most general sense, a large area, space based, NIR survey will provide
important input for the spectroscopic capabilities of the new
extremely large ground based telescopes and for the NGST. By
identifying quasars, galaxies, and GRBs to high redshift, SNAP will
set the table for OWL, CELT, and NGST in very much the same way that
the Palomar Sky Survey supported spectroscopy at 4m telescopes, and
the SDSS supports spectroscopy for 8--10m telescopes.

\vspace{0.1 in}

\section{NEAR INFRARED IMAGER CONCEPT}
\label{sect:nir_concept}

\begin{figure} [h]
   \begin{center}
   \begin{tabular}{c}
   \includegraphics [height=2.5in] {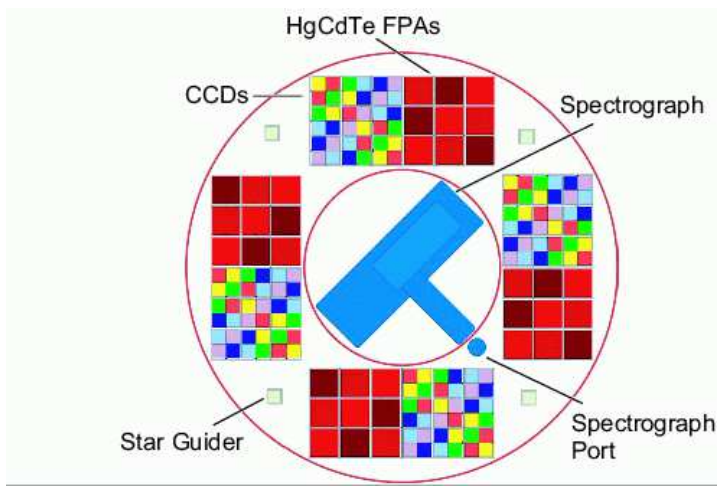}
   \end{tabular}
   \end{center}
   \caption[example] 
   { \label{fig:focal_plane} 
The SNAP near infrared imager concept:
The imager contains optical and IR sensors integrated on
a common focal plane. A total of thirty-six, 2k $\times$ 2k,
18 $\mu$m detectors will be used to cover the 900 nm to 1700
nm range. }
   \end{figure}

The NIR system for SNAP is an integral part of the overall focal
plane.  In the baseline concept, thirty-six 2k $\times$ 2k HgCdTe
imaging sensors with a pixel pitch of 18 $\mu$m will be
placed in four 3 $\times
$ 3 arrangements symmetric with the CCD placement as
shown in Fig.~\ref{fig:focal_plane}. Table 1 lists the current
performance specifications for the SNAP NIR system.  With over 150,000
pixels and a plate scale of 0.17$''$ per pixel, the NIR system will have
a field of view of 0.34 square degrees.  The HAWAII-2RG Focal Plane 
Array (FPA) manufactured by Rockwell Scientific Corporation (RSC)
is well matched to our
needs. These devices exhibit low read noise and dark current while
providing excellent quantum efficiency (typically 50\%-80\% 
over the wavelength
interval 1.0--1.7 $\mu$m).  A cutoff wavelength of $\lambda_c =
1.7$ $\mu$m is ideally suited to our mission since it renders our focal
plane effectively blind to the thermal
radiation from the warm telescope and allows passive cooling of 
the focal plane to 140 K. Three special filters fixed
above the FPAs in the pattern shown will provide overlapping
red-shifted $B$-band coverage from 0.9--1.7 $\mu$m. As SNAP repeatedly
steps across its target fields in the north and south ecliptic poles, every
supernova will be seen in every filter in both the visible and NIR.
Because of their larger linear size, each NIR filter will be visited
with twice the exposure time of the visible filters.  This, combined with the
time-dilated light curve, will ensure that Type Ia supernov{\ae} out to 
$z = 1.7$ will be detected with a S/N $>$ 3 at least 2 magnitudes below
peak brightness. 

\begin{table}[t]
\caption{SNAP NIR Performance Specifications.}
\label{tab:nir_specs}
\begin{center}       
\begin{tabular}{|l|l|l|} 
\hline
\rule[-1ex]{0pt}{3.5ex}  PARAMETER & SPECIFICATION & REASONING \\
\hline
\hline
\rule[-1ex]{0pt}{3.5ex}  Field of View & $\sim$0.3 square degrees & match
CCD FoV to observe \\
\cline{1-2}
\rule[-1ex]{0pt}{3.5ex}  Plate Scale & 0.17 arcsec/pixel & every SN
in every color\\
\hline
\rule[-1ex]{0pt}{3.5ex}  Wavelength Coverage & 0.9 $\mu$m - 1.7 $\mu$m
& to observe restframe $V$ band out to $z=1.7$ \\
\hline
\rule[-1ex]{0pt}{3.5ex}  Read Noise & 5 e$^-$ (w/multiple reads) &
Assures that photometry is \\
\cline{1-2}
\rule[-1ex]{0pt}{3.5ex}  Dark Current & $<$ 0.1 e$^-$/pixel/sec &
zodiacal light limited \\
\hline
\rule[-1ex]{0pt}{3.5ex}  Quantum Efficiency & $>60\%$ & to achieve
adequate S/N at $z=1.7$ within \\
\rule[-1ex]{0pt}{3.5ex}  & & time constraints\\
\hline
\rule[-1ex]{0pt}{3.5ex}  Filters & three special filters & to obtain
redshifted $B$ bands over 0.9 to 1.7 $\mu$m \\
\hline 
\end{tabular}
\end{center}
\end{table}


The HAWAII-2RG manufactured by RSC is currently the largest available
NIR device produced by Molecular Beam Epitaxy (MBE){\cite{MBE}}. MBE
allows precise control of the HgCdTe deposition which should result in
improved quantum efficiency and more uniform pixel response. The
latter is important for achieving precise SN photometry. In the near
future, competitive FPAs are expected to become available that can meet
the SNAP performance specifications.


The read noise specification of 5e$^-$ is approximately half the
background zodiacal light in the SNAP fields for nominal SNAP exposure
times. This specification will
likely be the hardest to achieve. It assumes that the intrinsic read
noise for a single read will be 10e$^-$ and that 4 reads at the beginning
and end of an exposure will reduce this by a factor of
$\sqrt4$. This noise performance has been achieved for HAWAII
multiplexers mated to HgCdTe material with a cut-off of 2.5 $\mu$m or
greater. Recently, excessively high read noise of 25e$^-$ or more
has been reported by the Wide Field Camera 3 (WFC3)
group\cite{bob_hill} for 1.7 $\mu$m material hybridized to HAWAII 
multiplexers.  Efforts are underway at RSC to solve this problem and first
results are expected within the year. Depending on the outcome, the
SNAP readout strategy may need to be modified to accommodate read noise
in excess of expectations.


The current performance specification for the NIR detector dark
current of $<$ 0.1 e$^-$/sec/pix is approximately half the background
zodiacal light for a nominal 300 sec exposure.  The dark current
depends sensitively on $\lambda_c$ and the detector temperature. Dark
currents as low as 0.02 e$^-$/sec/pix at 150 K have been achieved
with 1.7 $\mu$m devices\cite{bob_hill}. It is reasonable to expect that the
dark current should decrease by a factor 6--7 for the SNAP operating
temperature of 140 K. Thus we are confident that the SNAP dark current
specifications are achievable and the margin on this specification may
help us accommodate excess read noise.


The SNAP NIR quantum efficiency specification of $>$ 60\% results from
the need to detect Type Ia supernov{\ae} at 2 magnitudes below peak luminosity
out to $z = 1.7$ within the time constraints of the SNAP observation
plan.  Fabrication of WFC3 devices by RSC has led to a substantial
improvement in quantum efficiency over the wavelength range 
1.0--1.7 $\mu$m. Quantum
efficiencies as large as 85\% from 1.4--1.6 $\mu$m falling to 56\%
at 1.0 $\mu$m have been reported\cite{bob_hill}. The WFC3 development
should naturally provide
the SNAP quantum efficiency solution.


Sensitivity variations on the scale of a single pixel can 
significantly reduce photometric accuracy of undersampled images.
For SNAP, diffraction dominates the point-spread function at all NIR 
wavelengths. At 1 $\mu$m the Airy disk is Nyquist undersampled by a 
factor of three for a pixel size of 18 $\mu$m. Intrapixel variations
for PACE HgCdTe detectors have been measured by Gert Finger 
(ESO)\cite{Gert_Finger} and found to be quite large. The intrapixel response has not
been measured yet for MBE HgCdTe devices. These should 
have a more uniform pixel response because the internal fields can be
tuned to collect charge more efficiently from the edges of the pixels. 
Simulations\cite{Bernstein} show
that a simple 2$\times$2 dither pattern can reduce this component of
the photometry error to below 1\% for all wavelengths and for
reasonable levels of intrapixel variation. 

A program has begun to determine the science driven requirements
for the SNAP NIR system.  These requirements will be used to establish
SNAP science-grade specifications for the NIR FPAs.  Measurements
focusing on read noise, intrapixel variation, dark current and 
quantum efficiency will be made with the goal of demonstrating a SNAP
science-grade prototype meeting all SNAP science-grade specifications
within two years. 

\section{CONCLUSION}

Near infrared  observations in space are essential to understanding 
the nature of the dark energy that is causing the
expansion of the universe to accelerate. As an integral part of this
core mission, SNAP will collect NIR survey data of
unprecedented quality and scope, permitting detailed studies of weak
lensing and many secondary science objectives. 
The SNAP NIR system will be the largest NIR imager ever
deployed in space and will provide important input for targeting
future large ground and space-based telescopes.     
The NIR system concept that has been developed for SNAP is capable of
meeting this broad range of science goals.
A R\&D program is under way to firmly establish the science driven requirements
for the NIR system and to demonstrate that detectors can be obtained
that meet these requirements.

%
\appendix    

\acknowledgments     
 
This work was supported by the Director, Office of Science, of the
U.S.~Deparment of Energy under grant No.~DE-FG02-95ER40899 and contract 
No.~DE-AC03-76SF00098.  

%
%
%
%
%
%
%
%
%
%
%
%
%
%
%
\bibliography{spie_nir}   
\bibliographystyle{spiebib}   

\end{document}